\documentclass{article}

\usepackage[T1]{fontenc}
\usepackage[utf8]{inputenc}
\usepackage{hyperref}
\hypersetup{colorlinks,%
citecolor=blue,%
filecolor=blue,%
linkcolor=blue,%
urlcolor=blue,%
pdftex}

\title{MatD$^3$: A Database and Online Presentation Package for Research Data Supporting Materials Discovery, Design, and Dissemination}

\author{R. Laasner$^1$, X. Du$^1$, A. Tanikanti$^2$, C. Clayton$^3$, M. Govoni$^{2,4}$,\\
  G. Galli$^{2,4,5}$, M. Ropo$^6$, and V. Blum$^{1,7}$}

\date{%
  $^1$Department of Mechanical Engineering and Materials Science, Duke University,\\
  $^2$Pritzker School of Molecular Engineering, University of Chicago,\\
  $^3$Carnegie Mellon University,\\
  $^4$Materials Science Division and Center for Molecular Engineering, Argonne National Laboratory,\\
  $^5$Department of Chemistry, University of Chicago,\\
  $^6$Department of Physics, University of Turku,\\
  $^7$Department of Chemistry, Duke University}

\begin{document}

\maketitle

\section*{Summary}

The discovery of new materials as well as the determination of a vast set of materials properties for science and technology is a fast growing field of research, with contributions from many groups worldwide. Materials data from individual research groups is traditionally disseminated by means of loosely interconnected, peer-reviewed publications. Several data-centric efforts such as the Materials Project \cite{jain13}, Nomad \cite{nomad}, Aflow \cite{curtarolo12}, The Open Quantum Materials Database \cite{saal13}, or the Materials Data Facility \cite{blaiszik16} are making large segments of materials data publicly available, using a central repository. Because of the complexity and heterogeneity of materials science data, and the difficulty, in some cases, to assign given data to specific databases, researchers may find preferable to make data available in a distributed manner, on an individual basis, using locally hosted and controlled solutions. Of course distributed solutions do not exclude that data made available individually may also be contributed to large, centralized databases.

The package presented here (MatD$^3$) is intended to be a simple solution to make diverse datasets available individually and rapidly for reproducibility or other purposes. Furthermore, traditional data curation efforts such as SpringerMaterials \cite{springermaterials} or topical reviews have long supported the task of summarizing diverse sets of materials data from the community, but data in curated, reusable collections are not yet available for many classes of materials and material properties. Small and large research groups are struggling to find and access scientific data presented in scientific papers due to the lack of established queriable web interfaces to heterogenous sets of data.

MatD$^3$ is an open-source, dedicated database and web application framework designed to store, curate and disseminate experimental and theoretical materials data generated by individual research groups or research consortia. A research group can set up its own instance of MatD$^3$ and publish scientific results or simply use an existing online MatD$^3$ instance. Disseminating research data in this form enables broader access, reproducibility, and repurposing of scientific products. MatD$^3$ is a general purpose database that does not focus on any specific level of theory or experimental method. Instead, the focus is on storing and making accessible the data and making it straightforward to curate them.

MatD$^3$ deploys a structured (SQL) database which, along with RESTful APIs (application programming interface for representational state transfer), allows easy access and manipulation of the underlying datasets. The data submission interface is designed to be simple and intuitive, with numerous help tips to guide the user through each step. Setting up the servers includes fairly standard steps which are described in the online documentation \url{https://hybrid3-database.readthedocs.io/en/latest/}.

A counterpart to MatD$^3$ is the Qresp web application \cite{govoni19}, which is a tool to facilitate scientific data reproducibility by making available, in a distributed manner, all data and procedures presented in scientific papers, together with metadata to render them searchable and discoverable. The two main components of Qresp are \textit{i)} the curator, guiding users in the creation of metadata for the data that accompanies a publishable scientific work, and \textit{ii)} the explorer, a GUI for accessing datasets, exploring workflows, and downloading curated data, published in scientific papers.

We have implemented extensions to the Qresp and MatD$^3$ interfaces allowing collaborators to enter both their data as well as the associated publication and reproducibility information in a single workflow, thus reducing as much as possible the steps required to enter all research data for a given material. Specifically, the developed web GUI facilitates: \textit{i)} the creation of a MatD$^3$ entry with simultaneous generation of the corresponding metadata for Qresp curation and exploration; \textit{ii)} the generation of Qresp metadata for an already existing MatD$^3$ entry, and \textit{iii)} the generation of MatD$^3$ entries from Qresp metadata. While the developed coupling between two softwares is not limited to specific servers, we showcase the concurrent use of MatD$^3$ and Qresp with two specific instances installed at Duke University (\url{https://materials.hybrid3.duke.edu/} and \url{https://qresp.hybrid3.duke.edu/}). Having MatD$^3$ and Qresp instances communicating with each other requires no additional server configuring and is determined by user input during runtime (such as a URL for a specific server). Moreover, Qresp can serve as a frontend to any database that has the same API as MatD$^3$ for submitting new data.

In summary, the MatD$^3$ software provides a convenient, flexible, and complete solution for a web facing materials database for individual research groups and consortia. The availability of this solution will enable the community to take a significant step forward in making available the vast trove of newly generated materials research data directly for validation and reuse, in a convenient form that is directly amendable to processing by computational analysis or visualization software as well as quantitative data comparison and reproduction.

\section*{Acknowledgements}

This work was financially supported by the NSF under Award No. DMR-1729297 and and a dedicated supplement, Award No. DMR-1841206. G.G., M.G. and A.T. acknowledge support by MICCoM, as part of the Computational Materials Sciences Program funded by the U.S. Department of Energy, Office of Science, Basic Energy Sciences, Materials Sciences and Engineering Division.

\end{document}